\documentclass[fleqn,usenatbib,useAMS]{mnras}
 \usepackage{graphicx}
 \usepackage{amsmath}
 \usepackage{amssymb}
 \usepackage[T1]{fontenc}
 \usepackage{ae,aecompl}
 \usepackage{txfonts}
 \usepackage{enumerate}

 \title[The radiants of observed hyperbolic objects]
       {Where the Solar system meets the solar neighbourhood: patterns in 
        the distribution of radiants of observed hyperbolic minor bodies}

 \author[C. de la Fuente Marcos, R. de la Fuente Marcos and S. J. Aarseth]
        {Carlos~de~la~Fuente~Marcos,$^{1}$\thanks{E-mail: nbplanet@ucm.es}
         Ra\'ul~de~la~Fuente Marcos$^{1}$
         and
         Sverre J. Aarseth$^2$ \\
         $^1$Universidad Complutense de Madrid,
             Ciudad Universitaria, E-28040 Madrid, Spain \\
         $^2$Institute of Astronomy, University of Cambridge,
             Madingley Road, Cambridge CB3 0HA, UK}
 \date{Accepted 2018 February 1.
       Received 2018 January 31;
       in original form 2017 December 1}
 \pubyear{2018}
 
\begin{document}
  \label{firstpage}
  \pagerange{\pageref{firstpage}--\pageref{lastpage}}
  \maketitle

  \begin{abstract}
     Observed hyperbolic minor bodies might have an interstellar origin, but 
     they can be natives of the Solar system as well. Fly-bys with the known 
     planets or the Sun may result in the hyperbolic ejection of an originally 
     bound minor body; in addition, members of the Oort cloud could be forced 
     to follow inbound hyperbolic paths as a result of secular perturbations 
     induced by the Galactic disc or, less frequently, due to impulsive 
     interactions with passing stars. These four processes must leave 
     distinctive signatures in the distribution of radiants of observed 
     hyperbolic objects, both in terms of coordinates and velocity. Here, we 
     perform a systematic numerical exploration of the past orbital evolution 
     of known hyperbolic minor bodies using a full $N$-body approach and 
     statistical analyses to study their radiants. Our results confirm the 
     theoretical expectations that strong anisotropies are present in the 
     data. We also identify a statistically significant overdensity of 
     high-speed radiants towards the constellation of Gemini that could be 
     due to the closest and most recent known fly-by of a star to the Solar 
     system, that of the so-called Scholz's star. In addition to and besides 
     1I/2017~U1 (`Oumuamua), we single out eight candidate interstellar 
     comets based on their radiants' velocities.
  \end{abstract}


  \section{Introduction}
     The discovery \citep{2017MPEC....U..181B,2017MPEC....U..183M} and the subsequent study (see e.g. \citealt{2017ApJ...851L..38B,
     2017ApJ...850L..36J,2017ApJ...851L..31K,2017arXiv171009977M,2017Natur.552..378M,2017ApJ...851L...5Y,2018ApJ...852L...2B,2018NatAs...2..133F}) 
     of the first bona fide interstellar minor body, 1I/2017~U1 (`Oumuamua), has brought the subject of hyperbolic minor bodies into the 
     spotlight. Although some of the known ones might have had an interstellar origin like `Oumuamua, others (perhaps most of them) could be 
     natives of the Solar system. There are mechanisms capable of generating hyperbolic objects other than interstellar interlopers. They 
     include close encounters with the known planets or the Sun, for objects already traversing the Solar system inside the trans-Neptunian 
     belt; but also secular perturbations induced by the Galactic disc or impulsive interactions with passing stars, for more distant bodies 
     (see e.g. \citealt{2011A&A...535A..86F,2017A&A...604A..24F,2017MNRAS.472.4634K}). These last two processes have their sources beyond 
     the Solar system and may routinely affect members of the Oort cloud \citep{1950BAN....11...91O}, driving them into inbound hyperbolic 
     paths that may cross the inner Solar system, making them detectable from the Earth (see e.g. \citealt{1987Icar...69..185S}).
 
     Each and every object approaching from the outskirts of the Solar system appears to come from its own well-defined, unique location in 
     the sky, its radiant or antapex, and has a characteristic barycentric velocity that carries valuable information about its provenance. 
     The four processes listed above can induce strong anisotropies and leave distinctive signatures in the observed distribution of 
     radiants, both in terms of coordinates and velocity. The impact of some of these mechanisms on the perihelia of long-period comets has 
     been well studied (see e.g. \citealt{1996ApJ...472L..41M,1997CeMDA..69...77M,2002A&A...396..283D,2002MNRAS.335..641H}), but the 
     properties of the radiants of observed hyperbolic (eccentricity $>$1) minor bodies have never been studied in detail. Here, we carry 
     out a systematic numerical exploration of the past orbital evolution of known hyperbolic objects using a full $N$-body approach and 
     statistical analyses to study their radiants. This Letter is organized as follows. Section~2 introduces the tools used and the input 
     data. The distribution of radiants is presented and discussed in Section~3. The sample of internally produced hyperbolic minor bodies 
     is examined in Section~4. In Section~5, we study the population of former members of the Oort cloud. Candidate interstellar interlopers 
     are singled out in Section~6. Results are discussed and conclusions are summarized in Section~7.

  \section{Tools and input data}
     For minor bodies with very long orbital periods and extremely elongated orbits, the properties of their perihelia/aphelia encode a 
     significant amount of interesting dynamical information (see e.g. \citealt{2002MNRAS.335..641H}); for those currently following 
     hyperbolic paths, an equally relevant source of knowledge is in the radiant or point in the sky from which the incoming object appears 
     to originate. The analysis of the properties of the radiants of these interesting bodies can help in understanding their origin and 
     evolution. Aiming at extracting useful information ---namely, the positional and velocity distributions--- we have computed the 
     properties of the radiants associated with the orbit determinations available for these objects using full $N$-body simulations carried 
     out with a code written by \citet{2003gnbs.book.....A}\footnote{\url{http://www.ast.cam.ac.uk/~sverre/web/pages/nbody.htm}} that 
     implements a fourth-order version of the Hermite scheme described by \citet{1991ApJ...369..200M} without including any 
     non-gravitational forces. The model Solar system used in our calculations includes the perturbations from the eight major planets, with 
     the Earth--Moon system as two separate bodies. In addition, it incorporates the barycentre of the dwarf planet Pluto--Charon system and 
     the three most massive asteroids of the main belt. Positions and velocities in the barycentre of the Solar system for these bodies at 
     epoch JD 2458000.5 (2017-September-04.0 TDB, Barycentric Dynamical Time) have been provided by Jet Propulsion Laboratory's (JPL) 
     \textsc{horizons};\footnote{\url{https://ssd.jpl.nasa.gov/?horizons}} additional details are given in \citet{2012MNRAS.427..728D}. 
     Here, the present-day orbits of the known hyperbolic minor bodies ---339 with nominal heliocentric eccentricity $>$1, data as of 2018 
     January 18--- are integrated backwards for 0.1~Myr to compute the properties of their associated radiants. For these calculations, 
     we use input data provided by JPL's Solar System Dynamics Group Small-Body Database (SSDG SBDB, 
     \citealt{2015IAUGA..2256293G})\footnote{\url{https://ssd.jpl.nasa.gov/sbdb.cgi}} and the Minor Planets Center (MPC) Database 
     \citep{2016IAUS..318..265R}.\footnote{\url{http://www.minorplanetcenter.net/db_search}} As a reference and for a minor body moving with 
     an inbound velocity of 1~km~s$^{-1}$ ---i.e. it may travel 10\,000~au in less than 50\,000~yr--- that is the value of the escape 
     velocity at about 2\,000~au, our 0.1~Myr integrations back in time place such an object beyond 20\,000~au from the Sun, i.e. at the 
     outer Oort cloud (see e.g. \citealt{1981AJ.....86.1730H}). 

  \section{The distribution of radiants}
     The histograms presented in this section use a bin width computed using the Freedman-Diaconis rule \citep{FD81}, i.e. 
     $2\ {\rm IQR}\ n^{-1/3}$, where IQR is the interquartile range and $n$ is the number of data points. Averages, standard deviations, 
     medians, IQRs and other statistical parameters have been computed in the usual way (see e.g. \citealt{2012psa..book.....W}); we adopt
     Poisson statistics ($\sigma$=$\sqrt{n}$) to compute the error bars ---applying the approximation given by \citet{1986ApJ...303..336G} 
     when $n$$<$21, $\sigma$$\sim$$1+\sqrt{0.75+n}$.

\begin{figure}
        \centering
         \includegraphics[width=\linewidth]{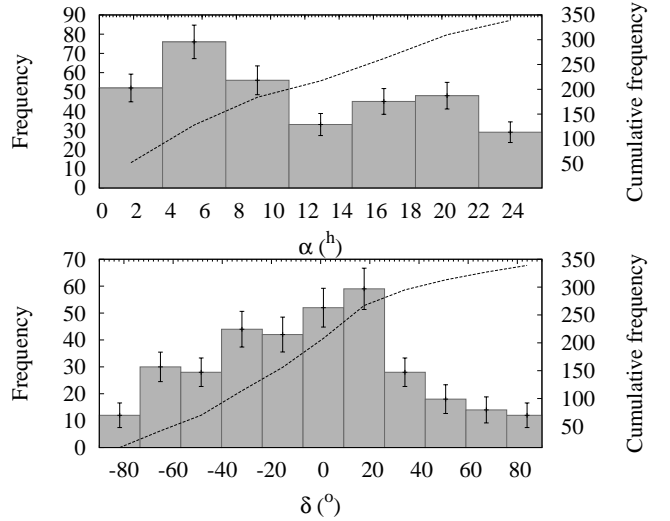}
         \caption{Distribution of the geocentric equatorial coordinates of the radiants of known hyperbolic minor bodies (nominal orbits); 
                  cumulative frequency in dashes, error bars from Poisson statistics (see Section~3).
                 }
         \label{coordinates}
      \end{figure}

     Fig.~\ref{coordinates} shows the distribution of geocentric equatorial coordinates of the radiants of known hyperbolic minor bodies 
     computed using the input data and the procedure described in the previous section. Here, the bin widths are 3\fh69 (top panel) and 
     16\fdg57 (bottom panel). The distribution in right ascension, $\alpha$, is somewhat asymmetric (top panel) with 193 radiants (out of 
     339) in the interval (0$^{\rm h}$, 12$^{\rm h}$). This is a 2.55$\sigma$ departure from an isotropic distribution, where 
     $\sigma$=$\sqrt{n}/2$ is the standard deviation for binomial statistics (see e.g. \citealt{2012psa..book.....W}). The presence of this 
     asymmetry might not be the result of observational bias because the radiant is not directly observed but computed once the orbit 
     determination is obtained. On the other hand, such asymmetry could be consistent with the one induced by a stellar passage through the 
     Oort cloud (see e.g. \citealt{2002A&A...396..283D}). The distribution in declination, $\delta$, is asymmetric as well (bottom panel), 
     but evenly distributed in terms of hemispheres as 176 radiants have southern declinations.
%
%-------------------------------------------------------------------------------------------------------------------------------------------
%
%
%-------------------------------------------------------------------------------------------------------------------------------------------
%

\begin{figure}
        \centering
         \includegraphics[width=\linewidth]{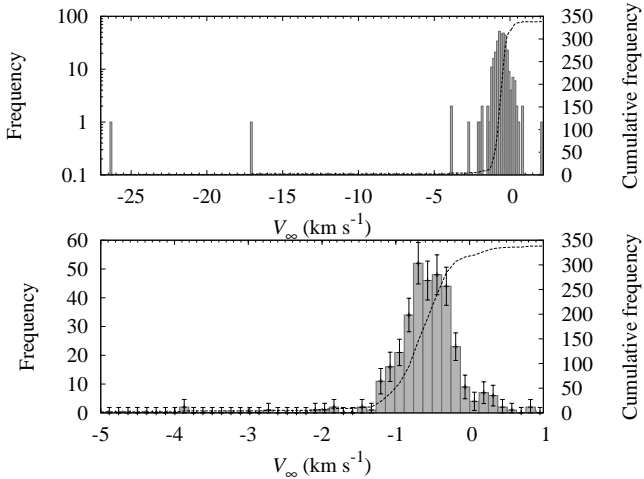}
         \caption{Distribution of the radiants' velocities of known hyperbolic minor bodies (cumulative frequency in dashes). The bottom
                 panel magnifies the section of the top one that includes most of the data.
                 }
         \label{velocity}
      \end{figure}

     Fig.~\ref{velocity} shows the radiant's velocity, $V_{\infty}$ (actually its proxy, the velocity at the end of the calculations),  
     histogram of the sample in Fig.~\ref{coordinates}; the bottom panel focuses on the bins with most of the entries, the bin width is 
     0.13~km~s$^{-1}$. The distribution is not Gaussian ---i.e. the average and standard deviation, $-$0.7$\pm$1.7~km~s$^{-1}$, cannot be 
     used to describe the velocity distribution adequately--- and includes a number of outliers (see Section 6). Out of 339 objects, 316 or 
     93.2 per cent show inbound (i.e. negative) barycentric velocities. The non-Gaussianity of the distribution suggests that multiple 
     processes may be shaping the observed velocity spread.
%
%-------------------------------------------------------------------------------------------------------------------------------------------
%
%
%-------------------------------------------------------------------------------------------------------------------------------------------
%

\begin{figure*}
        \centering
         \includegraphics[width=\linewidth]{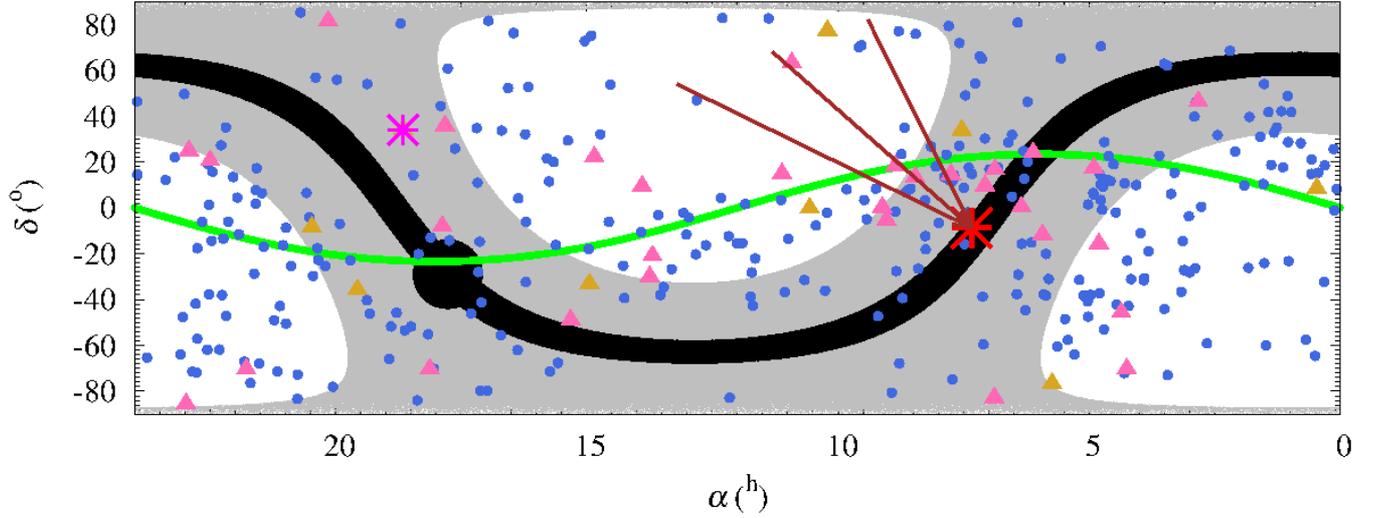}
         \caption{Distribution of radiants of known hyperbolic minor bodies in the sky. The radiant of 1I/2017~U1 (`Oumuamua) is represented 
                  by a pink star, those objects with radiant's velocity $>-1$~km~s$^{-1}$ are plotted as blue filled circles, the ones in 
                  the interval ($-1.5$, $-1.0$)~km~s$^{-1}$ are shown as pink triangles, and those $<$$-1.5$~km~s$^{-1}$ appear as goldenrod 
                  triangles. The current position of the binary star WISE J072003.20-084651.2, also known as Scholz's star, is represented 
                  by a red star, the convergent brown arrows represent its motion and uncertainty as computed by \citet{2015ApJ...800L..17M}. 
                  The ecliptic is plotted in green. The Galactic disc, which is arbitrarily defined as the region confined between Galactic 
                  latitude $-5${\degr} and 5\degr, is outlined in black, the position of the Galactic Centre is represented by a filled 
                  black circle; the region enclosed between Galactic latitude $-30${\degr} and 30\degr appears in grey. Data source: JPL's 
                  SSDG SBDB.
                 }
         \label{radiants}
      \end{figure*}

     Fig.~\ref{radiants} shows the distribution in equatorial coordinates of the radiants in Fig.~\ref{coordinates}. The distribution 
     exhibits a number of distinct concentrations that lead to the asymmetries present in Fig.~\ref{coordinates}. These clusters of radiants 
     are well away from that of 1I/2017~U1 (`Oumuamua), the pink star in Fig.~\ref{radiants}. The most obvious overdensity ---at 
     $\alpha$=7\fh4 and $\delta$=+16\fdg6--- may have as many as 36 radiants, or nearly 11 per cent of all the known ones, and about 22 per 
     cent (9/41) of the ones with radiant's velocity $<$$-1$~km~s$^{-1}$. Relevant comets in this group are C/1999~S4 (LINEAR), C/2007~W1 
     (Boattini), C/2010~X1 (Elenin), C/2012~S1 (ISON), or C/2013~A1 (Siding); some of them have experienced fragmentation/disintegration 
     events near perihelion. Other clusterings are observed towards $\alpha$=4\fh6 and $\delta$=+10\fdg0 ---14 possible members, relevant 
     comets are C/1956~F1-A (Wirtanen), C/1999~N4 (LINEAR), or C/2017~M4 (ATLAS)--- and $\alpha$=5\fh5 and $\delta$=$-$39\fdg0 ---16 
     possible members, relevant comets are C/1890~F1 (Brooks), C/2009~K5 (McNaught), or C/2013~G3 (PANSTARRS). Clusterings are also found in 
     the distribution of poles and perihelia of hyperbolic minor bodies \citep{2017RNAAS...1....5D}, but these can be due to observational 
     bias (particularly, the perihelion positions). 
%
%-------------------------------------------------------------------------------------------------------------------------------------------
%
%
%-------------------------------------------------------------------------------------------------------------------------------------------
%

\begin{figure}
        \centering
         \includegraphics[width=\linewidth]{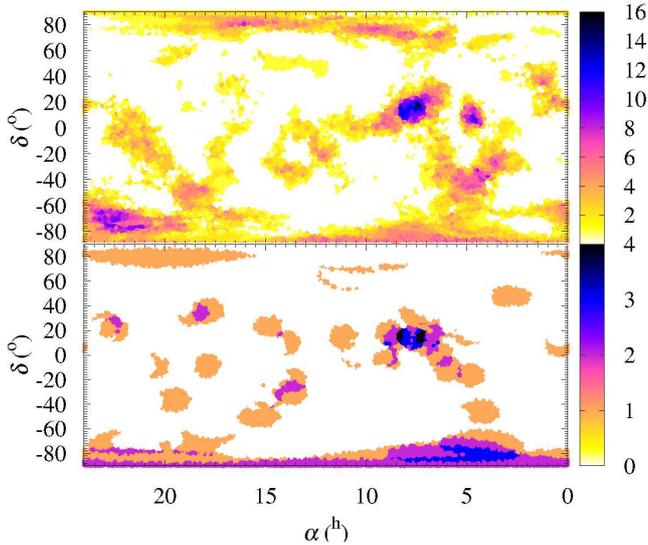}
         \caption{Statistical significance analysis of the distribution of radiants in Fig.~\ref{radiants}. Difference between counts from
                  a scan of the observed sample and that of an isotropic one; full sample analysis (top panel) and that of the subsample 
                  with velocity $<$$-1$~km~s$^{-1}$ (bottom panel). 
                 }
         \label{statsign}
      \end{figure}

     The distribution of radiants in Fig.~\ref{radiants} shows a number of conspicuous concentrations or overdensities both for the full 
     sample and for the subsample of objects with velocity $<$$-1$~km~s$^{-1}$ (plotted in pink and goldenrod); however, it is unclear from 
     the figure whether any of these overdensities are statistically significant. In order to make an informed decision, we have used a 
     population of hypothetical isotropic detections of the same size as reference. Such data set has been obtained by generating points 
     uniformly distributed on the surface of the celestial sphere using an algorithm due to \citet{MA72}. The positions of these points in 
     the sky have $\alpha\in(0^{\rm h}, 24^{\rm h})$ and $\delta\in(-90\degr, 90\degr)$. The excess of observed radiants with respect to a 
     uniformly distributed sample has been quantified by generating random points to cover the surface of the celestial sphere and counting 
     how many real and random radiants are found within 10{\degr} (our counting radius) of each random point. Our statistics is the 
     difference between real and uniformly distributed counts; as we are studying excesses not voids, negative differences are customarily 
     assigned a value of zero. Our experiments consider $2\times10^{5}$ random points (and counts) to scan the celestial sphere; multiple 
     random samples and several radii (5{\degr} and 15\degr) were tested to confirm that our overall results were neither affected by our 
     choice of counting radius nor by the actual random sample. Our analysis has been performed on the full sample ---Fig.~\ref{statsign}, 
     top panel, average difference of 1.3$\pm$1.9, median of 0, IQR=2--- and the subsample with velocity $<$$-1$~km~s$^{-1}$ 
     ---Fig.~\ref{statsign}, bottom panel, average difference of 0.3$\pm$0.6, median of 0, IQR=0. Fig.~\ref{statsign} confirms that the 
     overall spatial distribution of radiants of observed hyperbolic minor bodies is far from uniform and strongly anisotropic with several 
     statistically significant overdensities (up to 7.7$\sigma$) present in the data. The most relevant cluster of radiants is present in 
     both panels and it is located towards Gemini (see Section~5 for a detailed analysis). Other significant concentrations are observed 
     towards $\alpha$$\sim$$5^{\rm h}$, $\delta$$\sim$$+10\degr$ (top panel) and also $\alpha$=$3\fh3\pm0\fh3$, $\delta$=$-79\fdg3\pm0\fdg4$ 
     (bottom panel).  
%
%-------------------------------------------------------------------------------------------------------------------------------------------
%
%
%-------------------------------------------------------------------------------------------------------------------------------------------
%

  \section{Indigenously produced hyperbolics}
     Fig.~\ref{velocity} shows a tail of currently hyperbolic comets that, when integrated backwards, do not show inbound, i.e. negative, 
     velocities but outbound ones. In fact, our simulations show that these objects (about 10 per cent of the entire sample) were following 
     elliptical paths in the past, i.e. were bound to the Solar system, but they were ejected after experiencing close encounters with the 
     known planets and/or the Sun. Some comets that may have become hyperbolic in recent times could be C/1994~N1 (Nakamura-Nishimura-Machholz) 
     or C/2003~T4 (LINEAR). This might also be the case of comet C/1980~E1 (Bowell), which has a spectrum consistent with an origin in the 
     Solar system \citep{1982AJ.....87.1854J} and now has the second largest value of the eccentricity, $e$=1.0577 ($\sim$11\,500$\sigma$), 
     after that of 1I/2017~U1 (`Oumuamua), $e$=1.1995 ($\sim$1100$\sigma$). Such a high eccentricity might have been acquired after a fly-by 
     with Jupiter \citep{1982M&P....26..311B,2013RMxAA..49..111B}. Former interstellar comets, like 96P/Machholz~1 \citep{2007ApJ...664L.119L,
     2008AJ....136.2204S}, might also be eventually returned to deep space \citep{2015MNRAS.446.1867D}. The Solar system is also the source 
     of some artificially produced hyperbolic objects. Five spacecraft ---Pioneer~10 and 11, Voyager~1 and 2, and New Horizons--- currently 
     have outbound hyperbolic velocities with respect to the barycentre of the Solar system in excess of 10~km~s$^{-1}$ that will lead them 
     to deep space \citep{2016cosp...41E1273M}.

  \section{Oort cloud sources}
     The bulk of the distribution in Fig.~\ref{velocity} is probably compatible with the so-called Oort spike of new comets (see e.g. 
     \citealt{2017A&A...604A..24F,2017MNRAS.472.4634K}); in addition, the clusterings visible in Figs~\ref{radiants} and \ref{statsign} 
     could be consistent with some sort of weak comet shower coming from those directions (see e.g. \citealt{2002A&A...396..283D,2017A&A...604A..24F}).
     Figs~\ref{radiants} and \ref{statsign} show a statistically significant overdensity of hyperbolic comets with radiant inbound 
     velocities $>1$~km~s$^{-1}$ located towards the coordinates $\alpha=7^{\rm h}~25^{\rm m}~23^{\rm s}$, $\delta=+16\degr~38\arcmin~43\arcsec$ 
     (111\fdg3$\pm$0\fdg7, +16\fdg6$\pm$1\fdg1) in the constellation of Gemini. This location is well away from both Solar apex and 
     antapex. Most comets with radiants within the overdensity are widely regarded as new or Oort cloud comets. The presence of a coherent
     set of radiants hints at the outcome of a relatively recent stellar fly-by. Although the research on past and future close encounters
     between passing stars and the Solar system is still affected by significant uncertainties (see e.g. \citealt{2015A&A...575A..35B,
     2018A&A...609A...8B,2015MNRAS.449.2459D}) there is one case in which the dynamical parameters of the fly-by are relatively well 
     determined, that of the so-called Scholz's star \citep{2015ApJ...800L..17M} ---HIP~14473, other recent stellar fly-by, may have 
     approached within 0.22~pc, but 3.78~Myr ago \citep{2015MNRAS.449.2459D,2015MNRAS.454.3267F}. The current position of the binary star 
     WISE J072003.20-084651.2 \citep{2014A&A...561A.113S,2015AJ....149..104B} is plotted in Fig.~\ref{radiants} as a red star, the 
     convergent brown arrows represent its motion and uncertainty as computed by \citet{2015ApJ...800L..17M}. This low-mass binary may have 
     passed 52$^{+23}_{-14}$~kau from the Sun, 70$^{+15}_{-10}$~kyr ago; at its closest approach, it may have moved projected towards 
     $\alpha$=11\fh3$\pm$1\fh9 and $\delta$=+68\degr$\pm$14\degr (this area in Fig.~\ref{radiants} shows a relative void in the distribution 
     of radiants). It is difficult to attribute to mere chance the near coincidence in terms of timing and position in the sky between the 
     most recent known stellar fly-by and the statistically significant overdensity visible in Figs~\ref{radiants} and \ref{statsign}. It is 
     unclear whether other clusterings present may have the same origin or be the result of other, not-yet-documented, stellar fly-bys or 
     perhaps interactions with one or more unseen perturbers orbiting the Sun well beyond Neptune (see e.g. 
     \citealt{2014MNRAS.443L..59D,2014Natur.507..471T,2016AJ....151...22B}).

  \section{Interstellar interlopers}
     Fig.~\ref{velocity} shows a tail of hyperbolic minor bodies with inbound velocities well in excess of the median value of the radiant's
     velocity, $-0.57$~km~s$^{-1}$. In order to select candidates that may have an interstellar origin we adopt the cut-off value of 
     $-1.5$~km~s$^{-1}$; the difference between this value and the median is over twice the IQR, 0.44~km~s$^{-1}$, in absolute terms. Given 
     the distribution in Fig.~\ref{velocity}, we believe that any object with an inbound velocity $<$$-1.5$~km~s$^{-1}$ is a reasonably good 
     candidate to be an interstellar interloper ---the lower fence of Tukey's method \citep{1977eda..book.....T} to identify statistical 
     outliers is Q$_{1}-1.5$ IQR=$-1.45$~km~s$^{-1}$, where Q$_{1}$ is the lower quartile. Apart from 1I/2017~U1 (`Oumuamua), our list 
     includes C/1853 R1 (Bruhns), C/1997~P2 (Spacewatch), C/1999~U2 (SOHO), C/2002~A3 (LINEAR), C/2008~J4 (McNaught), C/2012~C2 (Bruenjes), 
     C/2012~S1 (ISON), and C/2017~D3 (ATLAS). Each candidate's radiant tends to be relatively well separated from the others (see 
     Fig.~\ref{radiants}), which suggests that they are dynamically uncorrelated. The best candidates are however C/2008~J4 (McNaught) and 
     C/2012~S1 (ISON) as their orbit determinations are more reliable than those of the others. In both cases, the inbound velocity is close 
     to $4$~km~s$^{-1}$ that is well away from that of the bulk of objects in Fig.~\ref{velocity}, bottom panel. Although C/1999~U2 might 
     have had $V_{\infty}$=$-$17.1~km~s$^{-1}$, the object might not be currently hyperbolic ---may now be captured as C/2005~W5 
     \citep{2005MPEC....Y...27K}--- but it was probably hyperbolic in the past. Interstellar interlopers could be the result of the 
     gravitational slingshot effect ---first discussed within the context of dense stellar systems by \citet{1974ApJ...190..253S}. The 
     prospect of detecting these bodies has been considered for decades (see e.g. \citealt{2016ApJ...825...51C,2017AJ....153..133E,
     2017ApJ...850L..38T}).
 
  \section{Discussion and conclusions}
     In sharp contrast to their bound counterparts and due to their unique nature, the orbital solutions of hyperbolic minor bodies are 
     based on relatively brief arcs of observation and this fact has an impact on their reliability. Our results depend on the quality of 
     the available orbit determinations, over 60 per cent of the solutions used here have uncertainties comparable or better than those 
     associated with that of 1I/2017~U1 (`Oumuamua) ---errors in $\alpha$, $\delta$, and $V_{\infty}$ are less than or close to 0\fh01, 
     0\fdg05, and 0.1~km~s$^{-1}$, respectively. This also applies to those objects being part of the overdensities of radiants identified 
     here. Out of 339 objects in the sample, 232 have reported uncertainties and 212 have eccentricity with statistical significance above 
     3$\sigma$ ---i.e. $(e-1)/\sigma>3$. Therefore, the overall conclusions of our investigation are expected to be essentially correct, but 
     those of some individual objects might not be. Regarding the statistical significance of the overdensities present in 
     Fig.~\ref{statsign}, the application of Tukey's method gives an upper fence value ---Q$_{3}+1.5$ IQR, where Q$_{3}$ is the upper 
     quartile--- for outliers of 5 for the top panel and 0 for the bottom panel, i.e. the overdensities are indeed statistically significant 
     according to Tukey's criterion. 

     In this Letter, we have explored the distribution of the radiants of observed hyperbolic minor bodies, both in terms of location in the 
     sky and kinematics. Our $N$-body calculations and subsequent statistical analyses lead to the following conclusions: 
     \begin{enumerate}[(i)]
        \item The distribution of the radiants of observed hyperbolic minor bodies is strongly anisotropic.
        \item Consistent with theoretical expectations, the distribution of radiants' velocities may result from the concurrent action of 
              four dynamical processes: local planetary (and Solar) fly-bys, external secular and impulsive perturbations on the Oort cloud, 
              and crossing paths with interstellar interlopers.  
        \item A statistically significant overdensity of hyperbolic comets with radiant's inbound velocity $>1$~km~s$^{-1}$ appears located 
              towards the coordinates $\alpha=7^{\rm h}~25^{\rm m}~23^{\rm s}$, $\delta=+16\degr~38\arcmin~43\arcsec$ (111\fdg3$\pm$0\fdg7, 
              +16\fdg6$\pm$1\fdg1) in the constellation of Gemini. 
        \item The overdensity of high-speed radiants appears to be consistent in terms of location and time constraints with the latest 
              known stellar fly-by, that of Scholz's star.
        \item Based on their current orbit determinations, eight hyperbolic comets emerge as good candidates to have an interstellar origin
              as they all have $V_{\infty}$$<$$-1.5$~km~s$^{-1}$: C/1853 R1 (Bruhns), C/1997~P2 (Spacewatch), C/1999~U2 (SOHO), C/2002~A3 
              (LINEAR), C/2008~J4 (McNaught), C/2012~C2 (Bruenjes), C/2012~S1 (ISON), and C/2017~D3 (ATLAS).
     \end{enumerate}

  \section*{Acknowledgements}
     We thank the anonymous referee for helpful comments and suggestions. CdlFM and RdlFM thank A.~I. G\'omez de Castro of the Universidad 
     Complutense de Madrid (UCM) for providing access to computing facilities. This work was partially supported by the Spanish `Ministerio 
     de Econom\'{\i}a y Competitividad' (MINECO) under grant ESP2014-54243-R. Part of the calculations and the data analysis were completed 
     on the EOLO cluster of the UCM, and we thank S. Cano Als\'ua for his help during this stage. EOLO, the HPC of Climate Change of the 
     International Campus of Excellence of Moncloa, is funded by the MECD and MICINN. In preparation of this Letter, we made use of the NASA 
     Astrophysics Data System, the ASTRO-PH e-print server, and the MPC data server.

  \bsp
  \label{lastpage}

\begin{thebibliography}{99}
     \bibitem[\protect\citeauthoryear{Aarseth}{2003}]{2003gnbs.book.....A} Aarseth S.~J., 2003,
             Gravitational N-body simulations.
             Cambridge Univ. Press, Cambridge, p.\ 27
     \bibitem[\protect\citeauthoryear{Bacci et al.}{2017}]{2017MPEC....U..181B} Bacci P. et al., 2017,
             MPEC Circ., MPEC 2017-U181
     \bibitem[\protect\citeauthoryear{Bailer-Jones}{2015}]{2015A&A...575A..35B} Bailer-Jones C.~A.~L., 2015, 
             A\&A, 575, A35
     \bibitem[\protect\citeauthoryear{Bailer-Jones}{2018}]{2018A&A...609A...8B} Bailer-Jones C.~A.~L., 2018, 
             A\&A, 609, A8
     \bibitem[\protect\citeauthoryear{Bannister et al.}{2017}]{2017ApJ...851L..38B} Bannister M.~T. et al., 2017, 
             ApJ, 851, L38
     \bibitem[\protect\citeauthoryear{Batygin \& Brown}{2016}]{2016AJ....151...22B} Batygin K., Brown M.~E., 2016,
             AJ, 151, 22
     \bibitem[\protect\citeauthoryear{Bolin et al.}{2018}]{2018ApJ...852L...2B} Bolin B.~T. et al., 2018, 
             ApJ, 852, L2
     \bibitem[\protect\citeauthoryear{Branham}{2013}]{2013RMxAA..49..111B} Branham R.~L., Jr., 2013, 
             RMxAA, 49, 111
     \bibitem[\protect\citeauthoryear{Buffoni, Scardia \& Manara}{1982}]{1982M&P....26..311B} Buffoni L., Scardia M., Manara A., 1982, 
             M\&P, 26, 311
     \bibitem[\protect\citeauthoryear{Burgasser et al.}{2015}]{2015AJ....149..104B} Burgasser A.~J. et al., 2015, 
             AJ, 149, 104
     \bibitem[\protect\citeauthoryear{Cook et al.}{2016}]{2016ApJ...825...51C} Cook N.~V., Ragozzine D., Granvik M., Stephens D.~C., 2016, 
             ApJ, 825, 51
     \bibitem[\protect\citeauthoryear{de la Fuente Marcos \& de la Fuente Marcos}{2012}]{2012MNRAS.427..728D} de la Fuente Marcos C.,
             de la Fuente Marcos R., 2012,
             MNRAS, 427, 728
     \bibitem[\protect\citeauthoryear{de la Fuente Marcos \& de la Fuente Marcos}{2014}]{2014MNRAS.443L..59D} de la Fuente Marcos C.,
             de la Fuente Marcos R., 2014,
             MNRAS, 443, L59
     \bibitem[\protect\citeauthoryear{de la Fuente Marcos \& de la Fuente Marcos}{2017}]{2017RNAAS...1....5D} de la Fuente Marcos C., 
             de la Fuente Marcos R., 2017, 
             Res. Notes AAS, 1, 5
     \bibitem[\protect\citeauthoryear{de la Fuente Marcos, de la Fuente Marcos \& Aarseth}{2015}]{2015MNRAS.446.1867D} de la Fuente Marcos C., 
             de la Fuente Marcos R., Aarseth S.~J., 2015, 
             MNRAS, 446, 1867
     \bibitem[\protect\citeauthoryear{Dybczy{\'n}ski}{2002}]{2002A&A...396..283D} Dybczy{\'n}ski P.~A., 2002, 
             A\&A, 396, 283
     \bibitem[\protect\citeauthoryear{Dybczy{\'n}ski \& Berski}{2015}]{2015MNRAS.449.2459D} Dybczy{\'n}ski P.~A., Berski F., 2015, 
             MNRAS, 449, 2459
     \bibitem[\protect\citeauthoryear{Engelhardt et al.}{2017}]{2017AJ....153..133E} Engelhardt T., Jedicke R., Vere{\v s} P., Fitzsimmons A., Denneau L., 
             Beshore E., Meinke B., 2017, 
             AJ, 153, 133
     \bibitem[\protect\citeauthoryear{Feng \& Bailer-Jones}{2015}]{2015MNRAS.454.3267F} Feng F., Bailer-Jones C.~A.~L., 2015, 
             MNRAS, 454, 3267
     \bibitem[\protect\citeauthoryear{Fitzsimmons et al.}{2018}]{2018NatAs...2..133F} Fitzsimmons A. et al., 2018, 
             Nature Astronomy, 2, 133
     \bibitem[\protect\citeauthoryear{Fouchard et al.}{2011}]{2011A&A...535A..86F} Fouchard M., Rickman H., Froeschl{\'e} C., Valsecchi G.~B., 2011, 
             A\&A, 535, A86
     \bibitem[\protect\citeauthoryear{Fouchard et al.}{2017}]{2017A&A...604A..24F} Fouchard M., Rickman H., Froeschl{\'e} C., Valsecchi G.~B., 2017, 
             A\&A, 604, A24
     \bibitem[\protect\citeauthoryear{Freedman \& Diaconis}{1981}]{FD81} Freedman D., Diaconis P., 1981,
             Z. Wahrscheinlichkeitstheor. Verwandte Geb., 57, 453
     \bibitem[\protect\citeauthoryear{Gehrels}{1986}]{1986ApJ...303..336G} Gehrels N., 1986, 
             ApJ, 303, 336
     \bibitem[\protect\citeauthoryear{Giorgini}{2015}]{2015IAUGA..2256293G} Giorgini J.~D., 2015,
             IAU General Assembly, Meeting \#29, 22, 2256293
     \bibitem[\protect\citeauthoryear{Hills}{1981}]{1981AJ.....86.1730H} Hills J.~G., 1981, 
             AJ, 86, 1730
     \bibitem[\protect\citeauthoryear{Horner \& Evans}{2002}]{2002MNRAS.335..641H} Horner J., Evans N.~W., 2002, 
             MNRAS, 335, 641
     \bibitem[\protect\citeauthoryear{Jewitt et al.}{1982}]{1982AJ.....87.1854J} Jewitt D.~C., Soifer B.~T., Neugebauer G., Matthews K., Danielson G.~E., 1982,
             AJ, 87, 1854
     \bibitem[\protect\citeauthoryear{Jewitt et al.}{2017}]{2017ApJ...850L..36J} Jewitt D., Luu J., Rajagopal J., Kotulla R., Ridgway S., Liu W., 
             Augusteijn T., 2017, 
             ApJ, 850, L36
     \bibitem[\protect\citeauthoryear{Knight et al.}{2017}]{2017ApJ...851L..31K} Knight M.~M., Protopapa S., Kelley M.~S.~P., Farnham T.~L., Bauer J.~M., 
             Bodewits D., Feaga L.~M., Sunshine J.~M., 2017, 
             ApJ, 851, L31 
     \bibitem[\protect\citeauthoryear{Kracht et al.}{2005}]{2005MPEC....Y...27K} Kracht R., Hammer D., Marsden B.~G., Sekanina Z., Chodas P., 2005, 
             MPEC Circ., MPEC 2005-Y27
     \bibitem[\protect\citeauthoryear{Kr{\'o}likowska \& Dybczy{\'n}ski}{2017}]{2017MNRAS.472.4634K} Kr{\'o}likowska M., Dybczy{\'n}ski P.~A., 2017, 
             MNRAS, 472, 4634
     \bibitem[\protect\citeauthoryear{Langland-Shula \& Smith}{2007}]{2007ApJ...664L.119L} Langland-Shula L.~E., Smith G.~H., 2007, 
             ApJ, 664, L119
     \bibitem[\protect\citeauthoryear{Makino}{1991}]{1991ApJ...369..200M} Makino J., 1991,
             ApJ, 369, 200
     \bibitem[\protect\citeauthoryear{Mamajek et al.}{2015}]{2015ApJ...800L..17M} Mamajek E.~E., Barenfeld S.~A., Ivanov V.~D., Kniazev A.~Y., 
             V{\"a}is{\"a}nen P., Beletsky Y., Boffin H.~M.~J., 2015, 
             ApJ, 800, L17
     \bibitem[\protect\citeauthoryear{Marsaglia}{1972}]{MA72} Marsaglia G., 1972,
             Ann. Math. Stat., 43, 645
     \bibitem[\protect\citeauthoryear{Masiero}{2017}]{2017arXiv171009977M} Masiero J., 2017, 
             ApJL, submitted (arXiv:1710.09977)
     \bibitem[\protect\citeauthoryear{Matese \& Whitmire}{1996}]{1996ApJ...472L..41M} Matese J., Whitmire D., 1996, 
             ApJ, 472, L41
     \bibitem[\protect\citeauthoryear{Matese, Whitman \& Whitmire}{1997}]{1997CeMDA..69...77M} Matese J.~J., Whitman P.~G., Whitmire D.~P., 1997, 
             Celest. Mech. Dyn. Astron., 69, 77
     \bibitem[\protect\citeauthoryear{McNutt \& Zurbuchen}{2016}]{2016cosp...41E1273M} McNutt R., Zurbuchen T.~H., 2016, 
             41st COSPAR Scientific Assembly, Abstract D1.1-15-16
     \bibitem[\protect\citeauthoryear{Meech et al.}{2017a}]{2017MPEC....U..183M} Meech K. et al., 2017a,
             MPEC Circ., MPEC 2017-U183
     \bibitem[\protect\citeauthoryear{Meech et al.}{2017b}]{2017Natur.552..378M} Meech K.~J. et al., 2017b, 
             Nature, 552, 378
     \bibitem[\protect\citeauthoryear{Oort}{1950}]{1950BAN....11...91O} Oort J.~H., 1950, 
             BAN, 11, 91
     \bibitem[\protect\citeauthoryear{Rudenko}{2016}]{2016IAUS..318..265R} Rudenko M., 2016,
             in Chesley S.~R., Morbidelli A., Jedicke R., Farnocchia D., eds,
             IAU Symp. 318: Asteroids: New Observations, New Models.
             Cambridge Univ. Press, Cambridge, p.\ 265
     \bibitem[\protect\citeauthoryear{Saslaw, Valtonen \& Aarseth}{1974}]{1974ApJ...190..253S} Saslaw W.~C., Valtonen M.~J., Aarseth S.~J., 1974, 
             ApJ, 190, 253
     \bibitem[\protect\citeauthoryear{Schleicher}{2008}]{2008AJ....136.2204S} Schleicher D.~G., 2008, 
             AJ, 136, 2204
     \bibitem[\protect\citeauthoryear{Scholz}{2014}]{2014A&A...561A.113S} Scholz R.-D., 2014, 
             A\&A, 561, A113
     \bibitem[\protect\citeauthoryear{Stern}{1987}]{1987Icar...69..185S} Stern A., 1987, 
             Icarus, 69, 185
     \bibitem[\protect\citeauthoryear{Trilling et al.}{2017}]{2017ApJ...850L..38T} Trilling D.~E. et al., 2017,
             ApJ, 850, L38
     \bibitem[\protect\citeauthoryear{Trujillo \& Sheppard}{2014}]{2014Natur.507..471T} Trujillo C.~A., Sheppard S.~S., 2014,
             Nature, 507, 471
     \bibitem[\protect\citeauthoryear{Tukey}{1977}]{1977eda..book.....T} Tukey J.~W., 1977, 
             Exploratory Data Analysis.
             Addison-Wesley, Reading, MA
     \bibitem[\protect\citeauthoryear{Wall \& Jenkins}{2012}]{2012psa..book.....W} Wall J.~V., Jenkins C.~R., 2012,
             Practical Statistics for Astronomers.
             Cambridge Univ. Press, Cambridge
     \bibitem[\protect\citeauthoryear{Ye et al.}{2017}]{2017ApJ...851L...5Y} Ye Q.-Z., Zhang Q., Kelley M.~S.~P., Brown P.~G., 2017, 
             ApJ, 851, L5
  \end{thebibliography}
\end{document}